# An Integrated-Photonics Optical-Frequency Synthesizer

Daryl T. Spencer[1*], Tara Drake[1], Travis C. Briles[1], Jordan Stone[1], Laura C. Sinclair[1], Connor Fredrick[1], Qing Li[2], Daron Westly[2], B. Robert Ilic[2], Aaron Bluestone[3], Nicolas Volet[3], Tin Komljenovic[3], Lin Chang[3], Seung Hoon Lee[4], Dong Yoon Oh[4], Myoung-Gyun Suh[4], Ki Youl Yang[4], Martin H. P. Pfeiffer[5], Tobias J. Kippenberg[5], Erik Norberg[6], Luke Theogarajan[3], Kerry Vahala[4], Nathan R. Newbury[1], Kartik Srinivasan[2], John E. Bowers[3], Scott A. Diddams[1], Scott B. Papp[1]

[1]*Time and Frequency Division, National Institute of Standards and Technology, Boulder, CO 80305 USA*
[2]*Center for Nanoscale Science and Technology, National Institute of Standards and Technology, Gaithersburg, MD 20899 USA*
[3]*University of California Santa Barbara, Santa Barbara, CA 93106 USA*
[4]*California Institute of Technology, Pasadena, CA 91125 USA*
[5]*Ecole Polytechnique Federale de Lausanne, Lausanne, Switzerland*
[6]*Aurrion Inc., Goleta, CA 93117 USA*
email: daryl.spencer@nist.gov

**Integrated-photonics microchips now enable a range of advanced functionalities for high-coherence applications such as data transmission[1], highly optimized physical sensors[2], and harnessing quantum states[3], but with cost, efficiency, and portability much beyond tabletop experiments. Through high-volume semiconductor processing built around advanced materials there exists an opportunity for integrated devices to impact applications cutting across disciplines of basic science and technology. Here we show how to synthesize the absolute frequency of a lightwave signal, using integrated photonics to implement lasers, system interconnects, and nonlinear frequency comb generation. The laser frequency output of our synthesizer is programmed by a microwave clock across 4 THz near 1550 nm with 1 Hz resolution and traceability to the SI second. This is accomplished with a heterogeneously integrated III/V-Si tunable laser, which is guided by dual dissipative-Kerr-soliton frequency combs fabricated on silicon chips. Through out-of-loop measurements of the phase-coherent, microwave-to-optical link, we verify that the fractional-frequency instability of the integrated-photonics synthesizer matches the $7.0\times10^{-13}$ reference-clock instability for a 1 second acquisition, and constrain any synthesis error to $7.7\times10^{-15}$ while stepping the synthesizer across the telecommunication C-band. Any application of an optical-frequency source would be enabled by the precision optical synthesis presented here. Building on the ubiquitous capability in the microwave domain, our results demonstrate a first path to synthesis with integrated photonics, leveraging low-cost, low-power, and compact features that will be critical for its widespread use.**

The electronics revolution that began in the mid-twentieth century was driven in part by advances related to the synthesis of radio and microwave frequency signals for applications in radar, navigation and communications systems. This formed a foundational aspect for more recent technologies of exceedingly wide impact, such as the global-positioning system (GPS) and cellular communications. Despite the ubiquity and importance of electronic synthesis, no comparable technology existed for electromagnetic signals in the optical domain until the introduction of the self-referenced optical frequency comb[4,5]. An optical frequency comb provides a phase-coherent link between microwave and optical domains, with an output consisting of an array of optical modes having frequencies given exactly by $v_n = n f_{rep} + f_{ceo}$, where $f_{rep}$ and $f_{ceo}$ are radio or microwave frequencies and $n$ is an integer. Over the past two decades, optical-frequency synthesizers employing frequency combs derived from mode-locked solid-state lasers have been demonstrated[6–8]. Such phase-coherent synthesis has proved invaluable for applications in frequency metrology and atomic timekeeping[9], microwave photonics[10], coherent light detection and ranging (LIDAR)[8], molecular spectroscopy[11], and astronomy[12], to name a few. Yet, while optical frequency comb technology has matured significantly over the past decade, such experiments are typically hand-assembled from optical components, occupy volumes ranging from (1 - 1000) liters, and require (0.1 - 1) kW of power.

A new opportunity for chip-integrated optical-frequency synthesis has emerged with broad development in heterogeneously integrated photonics[13] and the development microresonator frequency combs, or microcombs[14–24] across various platforms. Microresonators pumped by a continuous-wave (CW) laser generate a parametric four-wave mixing comb in dielectric media. Relying on waveguide confinement and high nonlinearity of the integrated photonics, microresonators provide a route to comb generation with only milliwatts of input power. Precise waveguide group-velocity dispersion (GVD) control,[25,26] combined with the discovery of low-noise dissipative Kerr solitons (DKS) in microcombs,[27–29] has led to ultra-broad optical spectra with dispersive waves[30,31] to enhance the signal-to-noise[32] in microcomb carrier-offset frequency detection[33–35]. In parallel, through heterogeneous integration it has become possible to seamlessly combine active and passive components, such as semiconductor lasers and amplifiers, electro-optic modulators, passive waveguides, photodiodes and CMOS electronics on a silicon-chip platform[13]. Our work utilizes the advantages of Kerr solitons and silicon photonics to realize an optical-frequency synthesizer, through absolute stabilization of the Kerr combs (Fig. 1a).

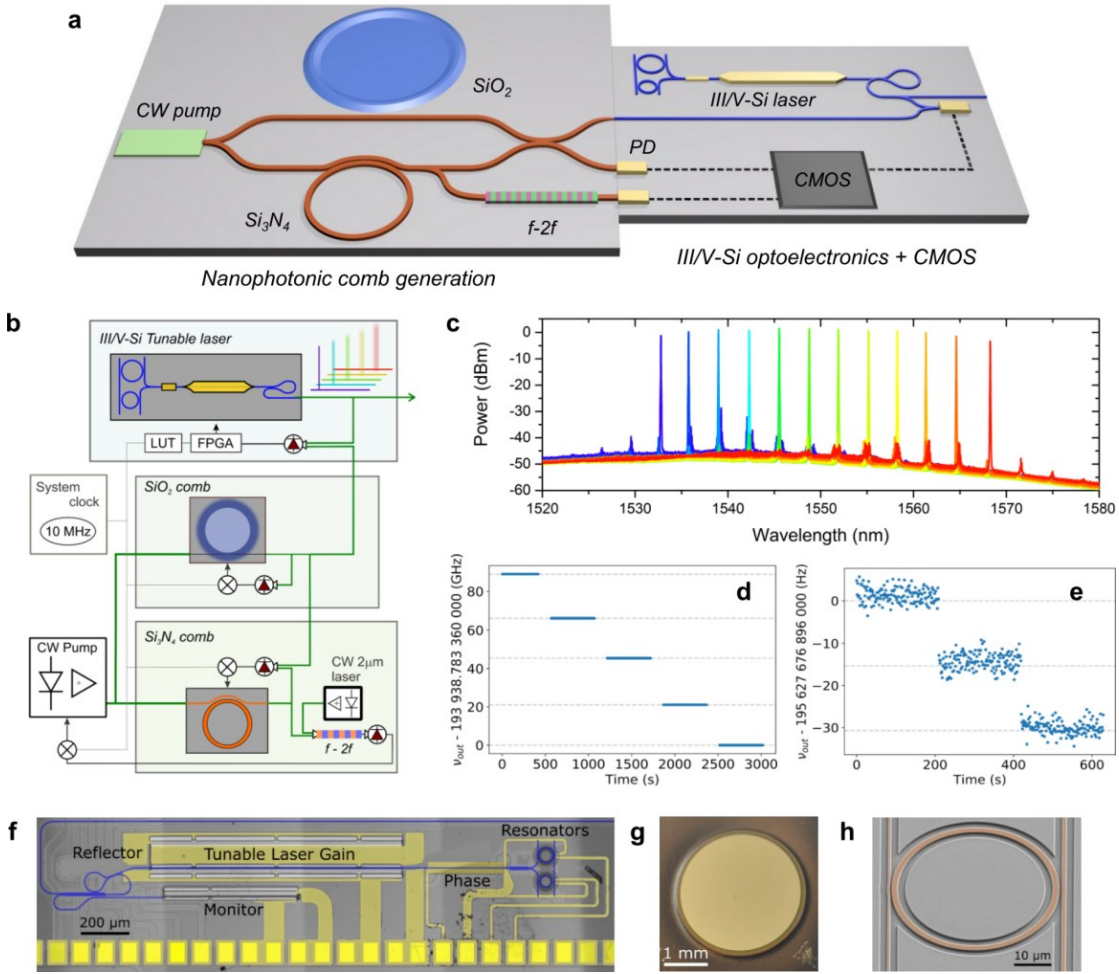

**Figure 1 | Accurate optical synthesis with an integrated laser and DKS dual-comb system.** (a) Concept of a integrated optical-frequency synthesizer with digital control and *f-2f* stabilization, utilizing the microcombs and tunable laser of this work. CMOS, complementary metal-oxide semiconductor, PD, photodetector. (b) To operate the integrated synthesizer, a 10 MHz clock is upconverted to 194 THz through dual-comb phase stabilization, and the tunable laser is programmed across its tuning range by phase locking to the stabilized combs, using a look-up-table (LUT) and field-programmable gate array (FPGA). (c) Optical spectra of the synthesized laser output at twelve discrete wavelengths across 32 nm. (d,e) Out-of-loop measurements of the synthesizer output as it is stepped. The data indicates the deviation between the integrated synthesizer and its setpoint when (d) mode-hopping across the 22 GHz $SiO_2$ modes, and (e) application of precise 15.36 Hz frequency steps. (f) SEM image of the heterogeneous III/V-Si tunable laser with false color electrodes (yellow) and waveguide layout (blue). (g) Photograph of the $SiO_2$ based wedge microresonator. (h) SEM image of the $Si_3N_4$ THz resonator with false color imposed on the waveguide regions.

Mirroring the framework of most traditional optical and microwave synthesizers, our system is composed of a tunable laser oscillator that we phase-lock to a stabilized microcomb reference. We use the C-band tunability, narrow linewidth, and rapid frequency control of a III/V-Si ring-resonator laser[36] (Fig. 1f) as the synthesizer output, and the phase-coherent microwave-to-optical connection of a fully stabilized DKS frequency comb. While an octave-bandwidth DKS comb is required for *f-2f* self-referencing, the flexibility of integrated photonics and DKS generation enables an interlocked *dual-comb* configuration that reduces optical power consumption by a factor of thirty. The DKS dual-comb consists of an octave-bandwidth, silicon-nitride comb with 1 THz mode spacing and a C-band spanning fused-silica comb with 22 GHz mode spacing. By phase stabilizing both comb spacings, $f_{rep,THz}$ and $f_{rep,GHz}$, respectively, and the silicon nitride comb's offset frequency, $f_{ceo,THz}$, we establish the precise factor of 19,403,904 phase-coherent multiplication from 10 MHz to the optical domain (Fig. 2a). With this tunable-laser and frequency comb system, we demonstrate synthesis across a 4 THz segment of the C-band by programming and dynamically stepping the output frequency. Since the role of any synthesizer is to output a phase-coherently multiplied version of the input clock, we characterize the integrated synthesizer primarily through its fluctuations with an out-of-loop frequency comb derived from the same clock. A fully integrated synthesizer would be a powerful tool for spectroscopy, sensing, optical communications, and other applications.

To demonstrate the integrated synthesizer, we carry out a series of experiments characterizing its output frequency. Standard spectrometer or interferometer measurements readily verify system performance at the MHz (or $10^{-8}$) level, similar to state-of-the-art instability in integrated-photonic references[37]. By measuring the integrated synthesizer with an auxiliary self-referenced, Erbium:fiber comb, we constrain the absolute frequency error between the output and the synthesizer's setpoint to <1.5 Hz. Beyond demonstration of the integrated-photonics architecture, the core result of our work is verification that the synthesizer offers sufficient phase control and synchronization in microwave-to-optical conversion (as do the auxiliary comb and our frequency counting electronics) to reveal a stable phase correlation between the CW output and the RF clock. Hence, the statistical fluctuations that lead to the synthesizer's instability, and our measurement of these, offer the complete description of the synthesizer's frequency performance.

The components of the integrated synthesizer, tunable laser and DKS dual-comb, and their key interconnections are shown in Fig. 1b. An external cavity pump laser is used to generate both the DKS combs, using independent control with single-sideband frequency shifters for each comb. First, an octave-spanning single pulse soliton is generated in a $Si_3N_4$ planar waveguide coupled resonator, shown in Fig. 1h, which also supports soliton-crystal configurations[38]. In addition to the anomalous GVD profile, waveguide-dispersion engineering creates dispersive wave peaks in optical power that appear at 999 nm and 2190 nm due to the zero-integrated GVD starting from the pump wavelength. With a radius of 23 μm, the threshold for octave-spanning spectra is brought to below 50 mW of on-chip pump power[32], at the expense of a $f_{rep,THz}$ of 1.014 THz that cannot be easily photodetected and reduced to a microwave frequency with conventional electronics. Rather, we rely on a second integrated-photonics frequency comb to bridge the gap between $Si_3N_4$ THz comb modes.

To do this, an $SiO_2$ wedge-based whispering-gallery-mode resonator with a quality factor of 180 million is used to create a DKS frequency comb at the $f_{rep,GHz} \approx$ 22 GHz[28] (Fig. 1g). This microwave repetition frequency is photodetected and phase locked to the RF clock, which represents the first step in the microwave-to-optical upconversion, here from $f_{clk}$=10 MHz to 22 GHz,

$$f_{rep,GHz} = 2197 \times f_{clk} \tag{1}$$

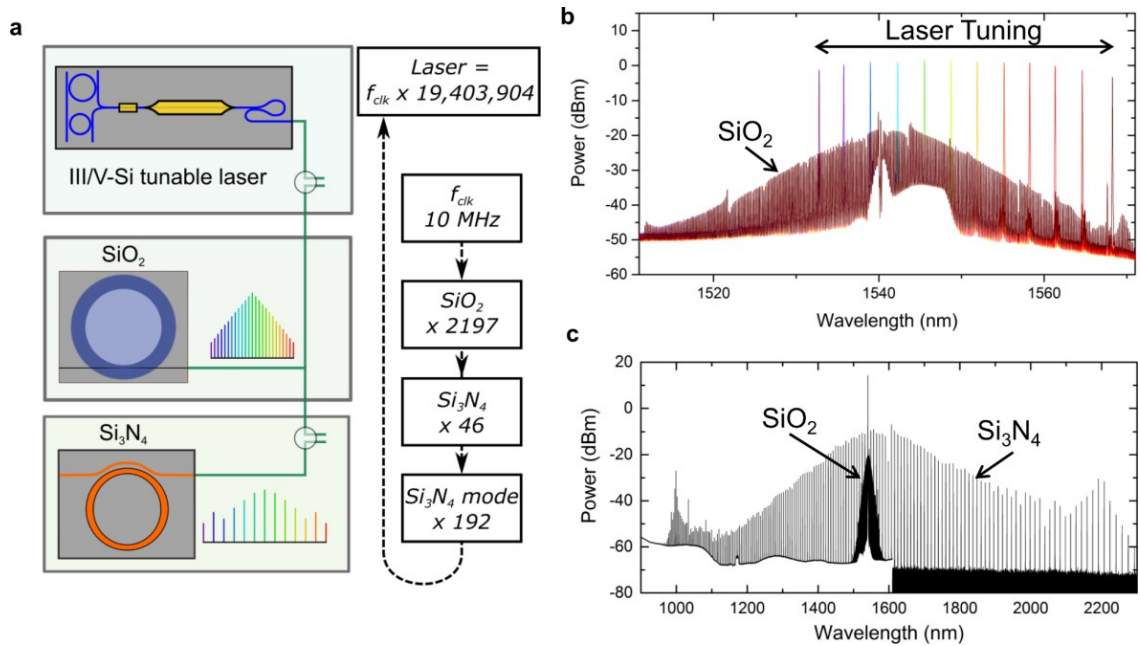

**Figure 2 | Optical spectra of the integrated devices.** (a) Schematic of spectral combination with the integrated devices, and the frequency chain used to multiply the 10 MHz clock to the optical domain. (b) Combined spectrum of the $SiO_2$ 22 GHz wedge resonator and the heterogeneously integrated III/V-Si tunable laser in the C-band. (c) Combined spectrum of the octave-spanning THz microcomb and the 22 GHz wedge microcomb.

The second step is detection of the 1.014 THz frequency spacing between $Si_3N_4$ comb teeth, which we accomplish using the 46th relative comb line from the $SiO_2$ comb. Operationally, we measure $f_{rep,THz}$ by detecting the optical heterodyne beat note between the two combs 1 THz away from the pump. We phase lock this signal to a synthesized radiofrequency, $f_1 = \alpha f_{clk}$ (where $\alpha$ is the ratio of two integers), after removing the relative contributions from the single-sideband frequency shifters and feeding back to the frequency of the $Si_3N_4$ pump laser[39,40]. Thus, we stabilize $f_{rep,THz}$ and transfer the $f_{clk}$ stability to 1.014 THz. The frequency of each of the $Si_3N_4$ THz comb lines with negative offset frequency and mode number N (N=192 at the pump) is then given by:

$$\begin{aligned} \nu_{THz} &= N \times f_{rep,THz} - f_{ceo,THz} \\ \nu_{THz,pump} &= 192 \times (46 \times f_{rep,GHz} + \alpha f_{clk}) - f_{ceo,THz}. \end{aligned} \tag{2}$$

Next, $f_{ceo,THz}$ locking is achieved by using the octave-spanning relationship of the THz lines at 1998 nm and dispersive wave peak at 999 nm (Fig. 2c). To aid *f-2f* self-referencing, an independent diode laser at 1998 nm is amplified up to 40 mW in a Thulium-doped fiber amplifier and frequency doubled in a waveguide PPLN device. Similar monolithic second-harmonic generation technologies have been demonstrated, and could be integrated with our system[41,42]. After detecting two heterodyne beats with the THz comb, $f_{999}$ and $f_{1998}$, each beat note is digitally divided by 64 and 32, respectively, and frequency mixing yields an $f_{ceo,THz}$ signal, $f_{ceo,THz}/64 = f_{999}/64 - f_{1998}/32$. Phase-locking this signal to a radiofrequency $f_2 = \beta f_{clk}$, through feedback to the $Si_3N_4$ pump power, completes the transfer of stability from $f_{clk}$ to all the THz comb lines spanning 130 THz to 300 THz.

The dual-stabilized combs serve as the backbone to guide the heterogeneously integrated III/V-Si tunable laser (Fig. 1f) for arbitrary optical frequency synthesis across the C-band (Fig. 1c). The tunable laser consists of InGaAsP multiple-quantum well epitaxial material that is wafer bonded onto a lithographically patterned silicon-on-insulator wafer[43]. Bias heaters integrated on the laser's Si based resonant reflectors and phase section are used to shift the lasing wavelength for initial alignment to the comb lines. By utilizing low loss Si waveguides relative to standard telecommunication-grade InP waveguide technology, reduced linewidths of ≈ 300 kHz are achieved. The combined optical spectrum of the $SiO_2$ comb and integrated laser's tuning range is shown in Fig. 2b. Heterodyning with the DKS dual-comb signal at a relative mode *m* from the pump creates a signal, $f_{beat}^{laser}$, for input to an FPGA based phased-locked-loop[44] (PLL, Fig. 3a) with a local oscillator of $f_3 = \gamma f_{clk}$, and digital division of 512. This final laser lock to the DKS dual-comb produces a fully stabilized, tunable synthesizer output, $\nu_{out}$.

$$\begin{aligned} \nu_{out} &= \nu_{THz,pump} + m \times f_{rep,GHz} + f_{beat}^{laser} \\ &= f_{clk}\left[192(46 \times 2197 + \alpha) + m \times 2197 - 64 \times \beta + 512 \times \gamma\right] \end{aligned} \tag{3}$$

This expression shows the output of our integrated-photonics synthesizer is uniquely and precisely defined relative to the input clock frequency in terms of user-chosen integers and ratios of integers ($\alpha, \beta, \gamma$).

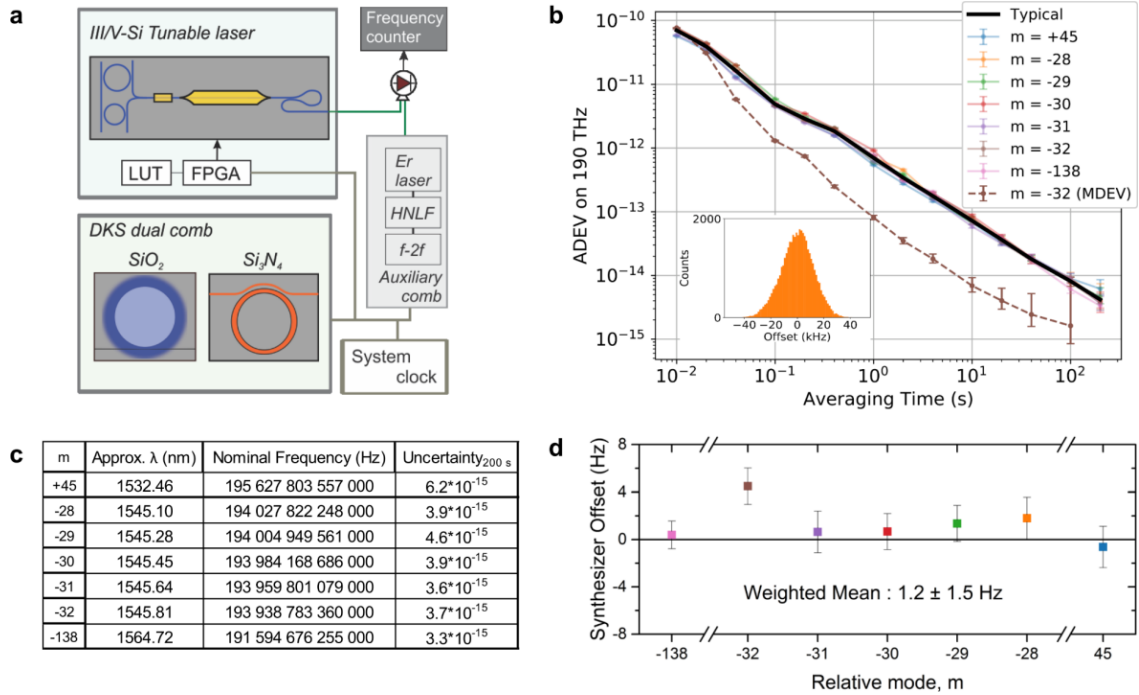

**Figure 3 | Stable optical synthesis with out-of-loop verification.** (a) Schematic of tunable laser locking, and frequency counting with the auxiliary comb. HNLF, highly nonlinear fiber. (b) Overlapping Allan deviation (ADEV) and modified Allan deviation (MDEV) analyses of the frequency-counter record from heterodyne detection with the auxiliary comb. In comparing many 10 ms counter-gate time acquisitions, the $1/\tau$ slope is consistent with a stable, phase-locked synthesizer, and the histograms of 500 seconds of data (inset for mode -28 only) show a Gaussian profile. 95% confidence intervals are derived using conservative flicker noise estimates[47]. (c) Table of nominal frequencies and instability as the synthesizer is stepped across the C-band. (d) Overview of the synthesizer frequency accuracy and precision. The ADEV at 100 seconds is used to estimate the uncertainty of the integrated synthesizer output, and the weighted mean is reported with a 95% confidence interval.

Out-of-loop frequency measurements are made by heterodyning $\nu_{out}$ against an auxiliary Erbium:fiber mode-locked laser frequency comb that is fully stabilized to the same $f_{clk}$. Agile tuning across SiO$_2$ comb lines (varying $m$) is shown in Fig. 1d and Hz-level tuning resolution on the same comb line (varying $\gamma$) is shown in Fig. 1e. Fig. 3 shows results from a study of the tunability and phase-locked operation of the synthesizer across all comb frequencies by locking to five adjacent SiO$_2$ comb lines, and to the highest and lowest wavelengths of the laser tuning range. Overlapping Allan deviation (ADEV) analysis of the counted beat notes against the auxiliary comb show the instability improving as $<10^{-12}/\tau$ for all recorded averaging times $\tau$, and an average instability of $(4.2 \pm 0.4) \times 10^{-15}$ at 200 seconds (Fig. 3b,c). More sophisticated triangular averaging analysis using the modified Allan deviation (MDEV) yields an order of magnitude better instability of $(9.2 \pm 1.4) \times 10^{-14}$ at 1 second. With this analysis, deviation from a $\tau^{(-3/2)}$ slope at longer averaging times shows unwanted flicker phase noise contributes to system performance. Still, the $1/\tau$ dependence of the ADEV data, which characterizes the fluctuations of the integrated synthesizer, indicates the stable phase relationship between the RF clock and the synthesized optical frequencies. From the mean values of the measured beats with the auxiliary comb, we can further analyze potential deviations of the synthesizer output from Eq. 3. Data compiled from seven such experiments is shown in Fig. 3d with 100 second ADEV error bars plotted, and the weighted mean of all data sets with a 95% confidence interval is $(1.2 \pm 1.5)$ Hz. Thus, based on these initial data we conclude that our integrated-photonics optical synthesizer accurately reproduces the input clock reference within an uncertainty of $7.7 \times 10^{-15}$.

To demonstrate the tunability of the optical frequency synthesizer, we perform two different types of laser tuning while locked to the stabilized comb system (Fig. 4). As a baseline, without changing the setpoint of the tunable laser phase lock, the raw data of the counted auxiliary comb beat note is shown in Fig. 4b after subtraction from the nominally expected frequency for 500 seconds. Then, we apply a bidirectional linear ramp over eight levels with a

2 second pause at each level to ensure successful locking (Fig. 4c). Finally, we program a series of setpoint frequencies to the FPGA PLL box to effectively write out the NIST logo (Fig. 4d). Excellent agreement is found between the expected offset frequencies and the counted beat note frequencies for all cases, illustrating good dynamic control of the synthesizer.

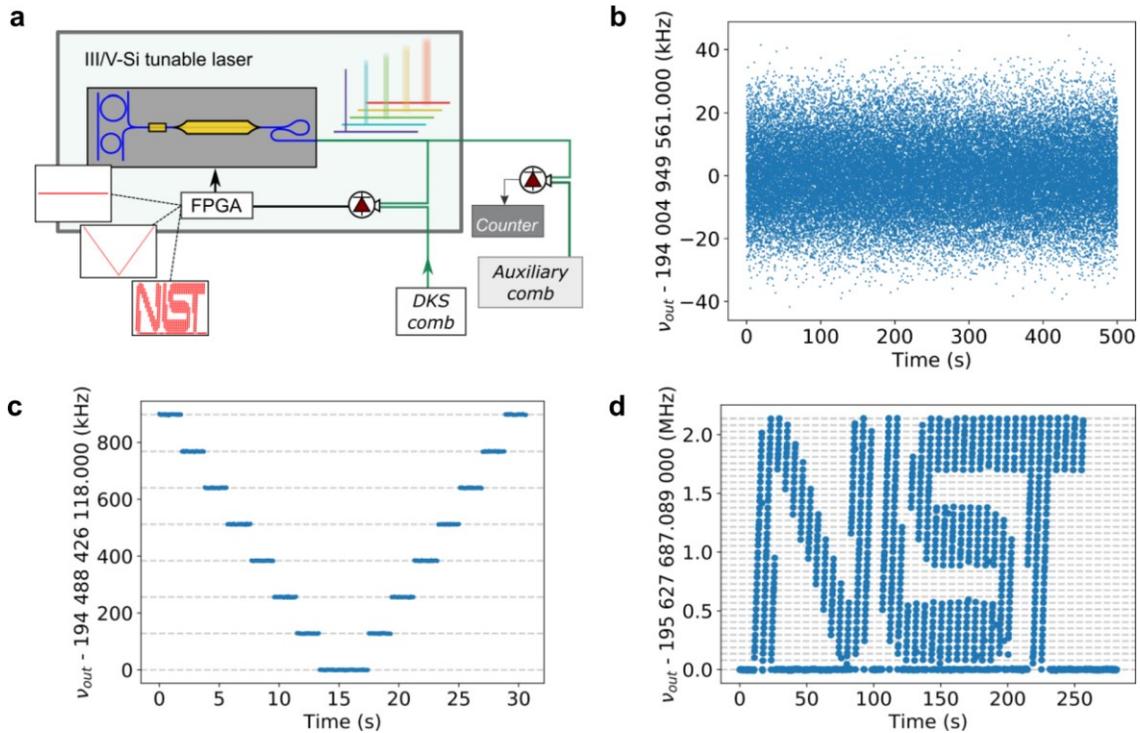

**Figure 4 | Arbitrary control of the integrated synthesizer.** (a) Step-wise control of the tunable-laser offset phase lock to the DKS comb and frequency counting. (b) Constant setpoint for 500 s at a 10 ms gate time. (c) Bidirectional linear ramp of the synthesizer via step control of the laser offset PLL setpoint (100 ms gate). (d) Arbitrary frequency control of the synthesizer across 40 frequency setpoints to write "NIST". A 30 ms gate time is used to oversample each frequency setpoint by 5 (150 ms pause/point) and every 5$^{th}$ datapoint is displayed.

In summary, the experiments we present with an optical-frequency synthesizer constructed from integrated photonics demonstrate that this technology has achieved the high precision and accuracy that formerly has been confined to tabletop mode-locked laser frequency comb devices. In particular, we report milestones that fully integrated tunable laser systems can support absolute optical-frequency stabilization, and that DKS frequency combs support full stabilization to implement a phase-coherent microwave-to-optical connection. With advancements in materials and fabrication techniques built around the nonlinear nanophotonics needed for DKS comb generation, it will be possible to support even more integration than has been possible in our current demonstration. The envisioned applications of such an integrated synthesizer device, based on the concepts reported in this paper, may most efficiently guide the route to implementation.

**Acknowledgements**
The authors thank Srico, Inc. for use of the waveguide PPLN device, Aurrion Inc. for use of the III/V-Si tunable laser, and Daniel Hickstein, David Carlson, and Zach Newman for providing comments on the paper. This research is supported by the Defense Advanced Research Projects Agency DODOS program and NIST. We thank Robert Lutwak and the DODOS program management team for helpful discussions throughout the experiment.

**Contributions**
D.T.S., T.D., T.C.B., and J.S. contributed equally to performing the system measurements and analyzing the experimental results. D.T.S., S.A.D., and S.B.P. prepared the manuscript. The integrated devices were fabricated and tested by Q.L., D.W., B.R.I., and K.S. (Si$_3$N$_4$), A.B., N.V., T.K., L.C., and E. N. (III/V-Si), and S.H.L., D.Y.O., M.S., K.Y.Y., and K.V. (SiO$_2$). N.V., L.C.S., C.F., M.H.P.F., and A.B. provided measurement support. T.J.K, E.N., K.V., K.S., N.R.N., L.T., J.E.B., S.A.D., and S.B.P. supervised and led the scientific collaboration.

**Methods**
**Device and experimental details**

The heterogeneously integrated III/V-Si device includes a tunable laser and a semiconductor optical amplifier (SOA). At room temperature, the laser emits up to ~4 mW CW power, and the SOA provides an on-chip small-signal gain >10 dB. The laser contains a gain section, a phase section, and two microresonators designed for high quality factor. The gain section and the SOA consist of electrically pumped InP-based quantum wells heterogeneously integrated on a Si waveguide[13]. Thermal heating of the Si microresonators and passive phase section is performed with current injection to metal heaters above the waveguides. By intentionally mismatching the microresonators' radii to utilize the Vernier effect, a narrowband intracavity optical filter selects the desired longitudinal mode for lasing with high side-mode-suppression-ratio[36]. Precise wavelength tuning and linewidth narrowing is performed by heating the phase section (Extended Data Fig. 1). Phase locking of the laser to the microcomb is performed by electronically dividing the beat note by 512 and utilizing FPGA based digital PLL+PI$^2$D feedback to the gain section of the laser. Other demonstrated works of high bandwidth optical-PLL phase locks to microcombs have also been demonstrated[45]. In the current system, the tunable laser can lock to either the SiO$_2$ or Si$_3$N$_4$ comb lines, which we have shown for m = -138, or 3 THz red of the pump laser.

A commercial external cavity diode laser is used as the shared pump for both microresonator comb generators, though we show that the III/V-Si tunable laser can be used to generate a low-noise solitons in the Si$_3$N$_4$ microresonator (Extended Data Fig. 2). The output of a 3 dB splitter goes to separate LiNbO$_3$ single-sideband modulators and Erbium-doped fiber amplifiers for each device. Frequency detuning from each microcomb resonance for soliton generation is controlled with an amplified voltage-controlled oscillator (VCO) and arbitrary waveform generator that produces a voltage ramp. Once initiated, feedback to each VCO controls $f_{rep,GHz}$ and $f_{rep,THz}$ for the appropriate device. Amplitude modulation on the Si$_3$N$_4$ microcomb to control $f_{ceo,THz}$ is performed with a free space acousto-optic modulator, though on-chip SOAs are expected to be viable as well. Lensed fibers with a 2.5 μm spot size are used to couple light on and off the Si$_3$N$_4$ chip with 7 dB of insertion loss per facet. During operation, the on-chip pump power for the Si$_3$N$_4$ microcomb is ≈ 160 mW, or ten times the threshold for soliton generation. Tapered single mode fiber is used to couple light to the SiO$_2$ microcomb, and ≈ 80 mW is launched in fiber for soliton generation at twelve times the soliton threshold. While the frequency shifter for the Si$_3$N$_4$ device can be ramped and held to an appropriate laser detuning for DKS, a sideband Pound-Drever-Hall lock is required after ramping to keep the SiO$_2$ pump frequency at the correct detuning. In a future implementation of our integrated synthesizer various technical improvements, such as improved on and off chip coupling and higher efficiency second-harmonic generation would make the 1998 nm diode laser unnecessary.

During current operation of the optical-frequency synthesizer, the separate single-sideband modulators for each microcomb device create a detectable offset in pump frequencies, ≈ 5 GHz in our experiment. This is readily subtracted from or added to the necessary heterodyne beat notes in the system using an electronic frequency mixer, mainly after $f_{rep,THz}$ detection between comb lines and after the III/V-Si laser heterodyne with the DKS comb. The calibrated gain sign of the tunable laser feedback loop insures that the tunable laser is on the appropriate side of the SiO$_2$ comb modes when electronically subtracting or adding this offset, and knowledge of the absolute difference in pump frequencies in not required for accurate optical-frequency synthesis. We observe non-zero synthesis error when the signal-to-noise ratio (SNR) of any heterodyne beat falls below the optimal level of 30 dB, but measurements reported here were acquired with sufficient SNR. We also observe and minimize contributions from out-of-loop optical and electrical path lengths, alignment drift, and glitches during long acquisitions.

**Auxiliary comb details and frequency counting**

The auxiliary comb used for out-of-loop verification of the optical-frequency synthesizer consists of a 250 MHz Erbium:fiber mode-locked laser frequency comb[46]. The laser output is amplified and spectrally broadened to an octave to enable self-referenced detection of the carrier envelope offset frequency, $f_{ceo}$. The 4$^{th}$ harmonic of $f_{rep}$ is phase-locked to a reference synthesizer at 999.999 544 MHz, and $f_{ceo}$ is electronically divided by 8 and phase locked to another synthesizer at 20 MHz. Both of these synthesizers are referenced to the same $f_{clk}$ that is the input to the integrated-photonics synthesizer, yielding a comb against which any frequency of the microcomb or tunable laser output can be compared.

The beat note frequency between the integrated-photonics synthesizer and the Erbium:fiber frequency comb is amplified and bandpass filtered (45 MHz bandwidth), after which a zero-dead-time Π frequency counter registers the frequency for a fixed gate time. The tunable laser PLL also contains an in-loop frequency counter, which showed tight phase locking of the laser to the microcomb at $<10^{-13}/\tau$, limited by the resolution of the counter. All RF synthesizers in the experimental setup, auxiliary comb, and frequency counter are tied to the same hydrogen maser signal, serving as $f_{clk}$.

**Data availability**

The raw data and analyzed datasets used in this study are available from the corresponding author on reasonable request.

**References**

1. Rumley, S. *et al.* Silicon Photonics for Exascale Systems. *J. Light. Technol.* **33,** 547–562 (2015).
2. Purdy, T. P., Grutter, K. E., Srinivasan, K. & Taylor, J. M. Quantum correlations from a room-temperature optomechanical cavity. *Science* **356,** 1265–1268 (2017).
3. O'Brien, J. L., Furusawa, A. & Vučković, J. Photonic quantum technologies. *Nat. Photonics* **3,** 687–695 (2010).
4. Hall, J. L. Nobel Lecture: Defining and measuring optical frequencies. *Rev. Mod. Phys.* **78,** 1279–1295 (2006).
5. Hansch, T. W. Nobel Lecture: Passion for precision. *Rev. Mod. Phys.* **78,** 1297–1309 (2006).
6. Jost, J., Hall, J. & Ye, J. Continuously tunable, precise, single frequency optical signal generator. *Opt. Express* **10,** 515–520 (2002).
7. Schibli, T. R. *et al.* Phase-locked widely tunable optical single-frequency generator based on a femtosecond comb. *Opt. Lett.* **30,** 2323–2325 (2005).
8. Giorgetta, F. R., Coddington, I., Baumann, E., Swann, W. C. & Newbury, N. R. Fast high-resolution spectroscopy of dynamic continuous-wave laser sources. *Nat. Photonics* **4,** 853–857 (2010).
9. Diddams, S. A. *et al.* An Optical Clock Based on a Single Trapped $^{199}Hg^+$ Ion. *Science* **293,** 825–828 (2001).
10. Xue, X. & Weiner, A. M. Microwave photonics connected with microresonator frequency combs. *Front. Optoelectron.* **9,** 238–248 (2016).
11. Thorpe, M. J., Moll, K. D., Jones, R. J., Safdi, B. & Ye, J. Broadband Cavity Ringdown Spectroscopy for Sensitive and Rapid Molecular Detection. *Science* **311,** 1595–1599 (2006).
12. Steinmetz, T. *et al.* Laser Frequency Combs for Astronomical Observations. *Science* **321,** 1335–1337 (2008).
13. Komljenovic, T. *et al.* Heterogeneous Silicon Photonic Integrated Circuits. *J. Light. Technol.* **34,** 20–35 (2016).
14. Del'Haye, P. *et al.* Optical frequency comb generation from a monolithic microresonator. *Nature* **450,** 1214–1217 (2007).
15. Savchenkov, A. A. *et al.* Tunable optical frequency comb with a crystalline whispering gallery mode resonator. *Phys. Rev. Lett.* **101,** 93902 (2008).
16. Razzari, L. *et al.* CMOS-compatible integrated optical hyper-parametric oscillator. *Nat. Photonics* **4,** 41–45 (2009).
17. Levy, J. S. *et al.* CMOS-compatible multiple-wavelength oscillator for on-chip optical interconnects. *Nat. Photonics* **4,** 37–40 (2009).
18. Grudinin, I. S., Yu, N. & Maleki, L. Generation of optical frequency combs with a $CaF_2$ resonator. *Opt Lett* **34,** 878–880 (2009).
19. Kippenberg, T. J., Holzwarth, R. & Diddams, S. A. Microresonator-based optical frequency combs. *Science* **332,** 555–559 (2011).
20. Papp, S. B. & Diddams, S. A. Spectral and temporal characterization of a fused-quartz-microresonator optical frequency comb. *Phys. Rev. A* **84,** 53833 (2011).
21. Ferdous, F. *et al.* Spectral line-by-line pulse shaping of on-chip microresonator frequency combs. *Nat. Photonics* **5,** 770–776 (2011).
22. Herr, T. *et al.* Universal formation dynamics and noise of Kerr-frequency combs in microresonators. *Nat. Photonics* **6,** 480–487 (2012).
23. Li, J., Lee, H., Chen, T. & Vahala, K. J. Low-pump-power, low-phase-noise, and microwave to millimeter-wave repetition rate operation in microcombs. *Phys. Rev. Lett.* **109,** 233901 (2012).
24. Huang, S.-W. *et al.* A broadband chip-scale optical frequency synthesizer at $2.7 \times 10^{-16}$ relative uncertainty.

*Sci. Adv.* **2,** e1501489 (2016).
25. Okawachi, Y. *et al.* Bandwidth shaping of microresonator-based frequency combs via dispersion engineering. *Opt. Lett.* **39,** 3535 (2014).
26. Yang, K. Y. *et al.* Broadband dispersion-engineered microresonator on a chip. *Nat. Photonics* **10,** 316–320 (2016).
27. Herr, T. *et al.* Temporal solitons in optical microresonators. *Nat. Photonics* **8,** 145–152 (2013).
28. Yi, X., Yang, Q.-F., Yang, K. Y., Suh, M.-G. & Vahala, K. Soliton frequency comb at microwave rates in a high-Q silica microresonator. *Optica* **2,** 1078–1085 (2015).
29. Coen, S., Randle, H. G., Sylvestre, T. & Erkintalo, M. Modeling of octave-spanning Kerr frequency combs using a generalized mean-field Lugiato-Lefever model. *Opt. Lett.* **38,** 37–39 (2013).
30. Brasch, V. *et al.* Photonic chip-based optical frequency comb using soliton Cherenkov radiation. *Science* **351,** 357–360 (2016).
31. Pfeiffer, M. H. P. *et al.* Octave-spanning dissipative Kerr soliton frequency combs in $Si_3N_4$ microresonators. *Optica* **4,** 684–691 (2017).
32. Li, Q. *et al.* Stably accessing octave-spanning microresonator frequency combs in the soliton regime. *Optica* **4,** 193–203 (2017).
33. Jost, J. D. *et al.* Counting the cycles of light using a self-referenced optical microresonator. *Optica* **2,** 706–711 (2015).
34. Del'Haye, P. *et al.* Phase-coherent microwave-to-optical link with a self-referenced microcomb. *Nat. Photonics* **10,** 516–520 (2016).
35. Brasch, V., Lucas, E., Jost, J. D., Geiselmann, M. & Kippenberg, T. J. Self-referencing of an on-chip soliton Kerr frequency comb without external broadening. *Light Sci. Appl.* **6,** 7–8 (2016).
36. Komljenovic, T. *et al.* Widely Tunable Narrow-Linewidth Monolithically Integrated External-Cavity Semiconductor Lasers. *IEEE J. Sel. Top. Quantum Electron.* **21,** 1501909 (2015).
37. Xiang, C. *et al.* Integrated chip-scale $Si_3N_4$ wavemeter with narrow free spectral range and high stability. *Opt. Lett.* **41,** 3309–3312 (2016).
38. Cole, D. C., Lamb, E. S., Del'Haye, P., Diddams, S. A. & Papp, S. B. Soliton crystals in Kerr resonators. *Arxiv:1610.00080* (2016).
39. Del'Haye, P., Arcizet, O., Schliesser, A., Holzwarth, R. & Kippenberg, T. Full Stabilization of a Microresonator-Based Optical Frequency Comb. *Phys. Rev. Lett.* **101,** 53903 (2008).
40. Papp, S. B. *et al.* Microresonator frequency comb optical clock. *Optica* **1,** 10–14 (2014).
41. Chang, L. *et al.* Thin film wavelength converters for photonic integrated circuits. *Optica* **3,** 531–535 (2016).
42. Guo, X., Zou, C. & Tang, H. Second-harmonic generation in aluminum nitride microrings with 2500 %/ W conversion efficiency. *Optica* **3,** 1126–1131 (2016).
43. Heck, M., Bauters, J., Davenport, M. & Doylend, J. Hybrid silicon photonic integrated circuit technology. *IEEE J. Sel. Top. Quantum Electron.* **19,** 6100117 (2013).
44. Sinclair, L. C. *et al.* A compact optically coherent fiber frequency comb. *Rev. Sci. Instrum.* **86,** 81301 (2015).
45. Arafin, S. *et al.* Power-Efficient Kerr Frequency Comb Based Tunable Optical Source. *IEEE Photonics J.* **9,** 6600814 (2017).
46. Ycas, G., Osterman, S. & Diddams, S. a. Generation of a 660–2100 nm laser frequency comb based on an erbium fiber laser. *Opt. Lett.* **37,** 2199–2201 (2012).
47. Greenhall, C. A. & Riley, W. J. Uncertainty of Stability Variances. *Proc. PTTI 2003* 267–280 (2003).

**Extended Data Figures**

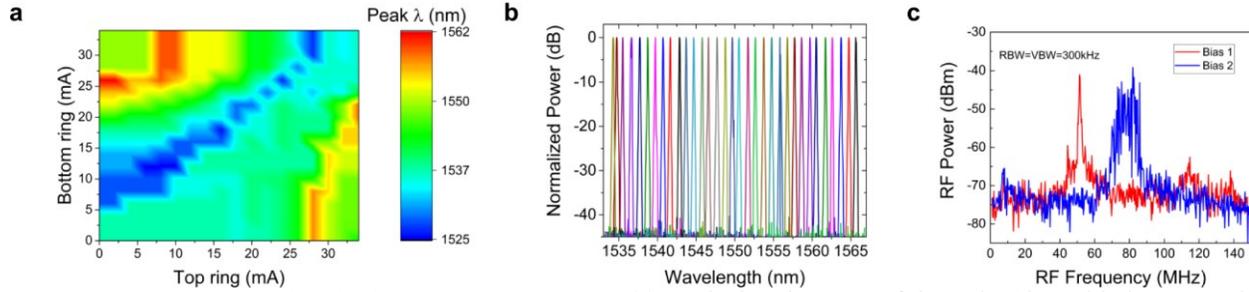

**Extended Data Figure 1 | III/V-Si laser tuning details.** (a) Typical tuning map of the III/V-Si tunable laser's peak wavelength versus current applied to each heater above the ring resonators. (b) Normalized optical spectra showing > 40 dB of side-mode suppression ratio across the tuning range. (c) Typical unlocked RF beat notes between the tunable laser and the auxiliary comb. Careful control of the phase section heater is required to reach all wavelengths in the tuning range and to reduce the linewidth of the laser (blue to red) to achieve best phase locking to the microcombs.

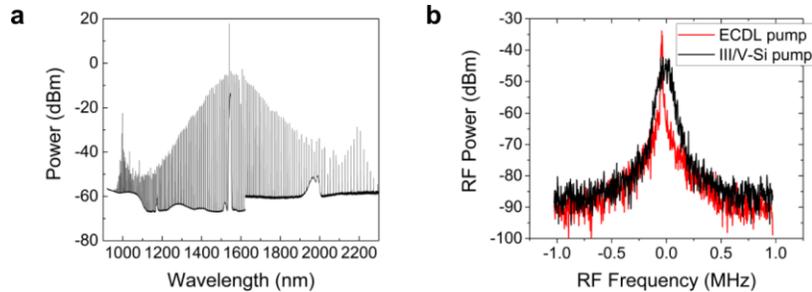

**Extended Data Figure 2 | Demonstration of pumping the $Si_3N_4$ THz microcomb with the III/V-Si laser** (a) Output optical spectrum of the THz microcomb showing dual-dispersive waves. (b) Comparison of electro-optic repetition rate detection[48] when using the same III/V-Si laser (black) and external cavity diode laser (ECDL, red) from the main experiment to pump the THz microcomb.

**Extended Data References**

48. Del'Haye, P., Papp, S. B. & Diddams, S. A. Hybrid electro-optically modulated microcombs. *Phys. Rev. Lett.* **109,** 263901 (2012).